\documentclass{llncs}
\usepackage{array}
\usepackage{color}
\usepackage{comment}
\usepackage{graphicx}
\usepackage[utf8]{inputenc}
\usepackage{wrapfig}

\author{Finn {\AA}rup Nielsen}
\title{A new ANEW: Evaluation of a word list for sentiment analysis in microblogs}

\authorrunning{Nielsen}
\tocauthor{Finn {\AA}rup Nielsen}
\institute{DTU Informatics, Technical University of Denmark, Lyngby,
  Denmark. 
  \email{fn@imm.dtu.dk}, \texttt{http://www.imm.dtu.dk/\~{}fn/}}

\begin{document}
\maketitle

\begin{abstract}
  Sentiment analysis of microblogs such as Twitter has recently gained
  a fair amount of attention.
  One of the simplest sentiment analysis approaches compares the words
  of a posting against a labeled word list, where each word has been
  scored for valence, --- a ``sentiment lexicon'' or ``affective word
  lists''. 
  There exist several affective word lists, e.g., ANEW (Affective
  Norms for English Words) developed before the advent of
  microblogging and sentiment analysis. 
  I wanted to examine how well ANEW and other word lists performs for the
  detection of sentiment strength in microblog posts in comparison
  with a new word list specifically constructed for microblogs. 
  I used manually labeled postings from Twitter scored for sentiment.
  Using a simple word matching I show that the new word list 
  may perform better than ANEW, though not as good as the more
  elaborate approach found in SentiStrength.
\end{abstract}

\section{Introduction}

Sentiment analysis has become popular in recent years.
Web services, such as socialmention.com, may even score microblog postings on 
Identi.ca and Twitter for sentiment in real-time.
One approach to sentiment analysis starts with labeled texts and uses
supervised machine learning trained on the labeled text data to classify
the polarity of new texts \cite{PangB2008Opinion}.
Another approach creates a sentiment lexicon and scores the text based
on some function that describes how the words and phrases of the text
matches the lexicon. 
This approach is, e.g., at the core of the \emph{SentiStrength}
algorithm \cite{ThelwallM2010Sentiment}.

It is unclear how the best way is to build a sentiment lexicon.
There exist several word lists labeled with emotional valence, e.g.,
ANEW \cite{BradleyM1999Affective}, General Inquirer, OpinionFinder
\cite{WilsonT2005Recognizing}, SentiWordNet and WordNet-Affect as well as the
word list included in the SentiStrength software
\cite{ThelwallM2010Sentiment}. 
These word lists differ by the words they include, e.g., some do not
include strong obscene words and Internet slang acronyms, such as
``WTF'' and ``LOL''.  
The inclusion of such terms could be important for reaching good
performance when working with short informal text found in Internet
fora and microblogs.  
Word lists may also differ in whether the words are scored with
sentiment strength or just positive/negative polarity.

I have begun to construct a new word list with sentiment strength and
the inclusion of Internet slang and obscene words.
Although we have used it for sentiment analysis on Twitter data
\cite{HansenL2011Good_accepted} we have not yet validated it.
Data sets with manually labeled texts can evaluate the performance of the
different sentiment analysis methods.
Researchers increasingly use Amazon Mechanical Turk (AMT) for creating 
labeled language data, see, e.g., \cite{AkkayaC2010Amazon}.
Here I take advantage of this approach.

\section{Construction of word list}

My new word list was initially set up in 2009 for tweets downloaded for
online sentiment analysis in relation to the United
Nation Climate Conference (COP15).
Since then it has been extended. 
The version termed AFINN-96 distributed on the
Internet\footnote{http://www2.imm.dtu.dk/pubdb/views/publication\_details.php?id=59819}
has 1468 different words, including a few phrases.
The newest version has 2477 unique words, including 15 phrases that
were not used for this study.
As SentiStrength\footnote{http://sentistrength.wlv.ac.uk/} it uses a
scoring range from $-5$ (very negative) to $+5$ (very positive).
For ease of labeling I only scored for valence, leaving out, e.g.,
subjectivity/objectivity, arousal and dominance.
The words were scored manually by the author.

The word list initiated from a set of obscene words
\cite{BaudhuinE1973Obscene,SapolskyB2008Rating} as well as a few
positive words. It was gradually extended by examining Twitter
postings collected for COP15 particularly the postings which scored high on
sentiment using the list as it grew.
I included words from the public domain \emph{Original Balanced Affective Word
List}\footnote{http://www.sci.sdsu.edu/CAL/wordlist/origwordlist.html}
by Greg Siegle. 
Later I added Internet slang by browsing the Urban
Dictionary\footnote{http://www.urbandictionary.com} including acronyms such
as WTF, LOL and ROFL.
The most recent additions come from the large word list by Steven
J. DeRose, \emph{The Compass DeRose Guide to Emotion
Words}.\footnote{http://www.derose.net/steve/resources/emotionwords/ewords.html} 
The words of DeRose are categorized but not scored for valence with
numerical values. 
Together with the DeRose words I browsed Wiktionary and the synonyms it
provided to further enhance the list.
In some cases I used Twitter to determine in
which contexts the word appeared.
I also used the Microsoft Web n-gram similarity Web service
(``Clustering words based on context
similarity''\footnote{http://web-ngram.research.microsoft.com/similarity/})
to discover relevant words. 
I do not distinguish between word categories so to avoid ambiguities
I excluded words such as patient, firm, mean, power and frank.
Words such as ``surprise''---with high arousal but with variable
sentiment---were not included in the word list. 

\begin{wrapfigure}[13]{r}{0.5\textwidth}
  \vspace{-0em}
  \centering
  \includegraphics[width=.5\textwidth]{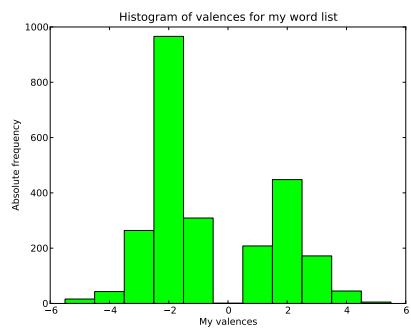}
  \caption{Histogram of my valences.}
  \label{fig:hist}
\end{wrapfigure}

Most of the positive words were labeled with +2 and most of the negative words with
--2, see the histogram in Figure~\ref{fig:hist}.
I typically rated strong obscene words, e.g., as listed in
\cite{BaudhuinE1973Obscene}, with either --4 or --5.
The word list have a bias towards negative words (1598, corresponding
to 65\%) compared to positive words (878). A single phrase was labeled
with valence 0. 
The bias corresponds closely to the bias found in the
OpinionFinder sentiment lexicon (4911 (64\%) negative and 2718
positive words). 

\begin{wrapfigure}[16]{r}{0.5\textwidth}
  \vspace{-2em}
  \centering
  \includegraphics[width=0.5\textwidth]{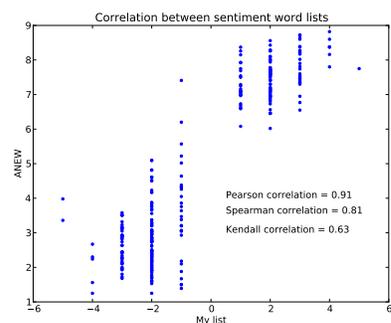}
  \caption{Correlation between ANEW and my new word list.}
  \label{fig:anew}
\end{wrapfigure}

I compared the score of each word with mean valence of ANEW.
Figure~\ref{fig:anew} shows a scatter plot for this comparison
yielding a Spearman's rank correlation on 0.81 when words are directly
matched and including words only in the intersection of the two word lists. 
I also tried to match entries in ANEW and my word list by applying
Porter word stemming (on both word lists) and WordNet lemmatization (on
my word list) as implemented in NLTK
\cite{BirdS2009Natural}. 
The results did not change significantly.

When splitting the ANEW at valence 5 and my list at valence 0 
I find a few discrepancies: aggressive, mischief, ennui, hard, silly,
alert, mischiefs, noisy.
Word stemming generates a few further discrepancies, e.g.,
alien/alienation, affection/affected, profit/profiteer. 

Apart from ANEW I also examined General Inquirer and the OpinionFinder
word lists.
As these word lists report polarity I associated words with
positive sentiment with the valence +1 and negative with --1.
I furthermore obtained the sentiment strength from SentiStrength via
its Web service\footnote{http://sentistrength.wlv.ac.uk/} and
converted its positive and negative sentiments to one single value by
selecting the one with the numerical largest value and zeroing the
sentiment if the positive and negative sentiment magnitudes were equal. 

\section{Twitter data}

For evaluating and comparing the word list with ANEW, General
Inquirer, OpinionFinder and SentiStrength a data set of 1,000 tweets
labeled with AMT was applied.  
These labeled tweets were collected by Alan Mislove for the
\emph{Twittermood}/``Pulse of a
Nation''\footnote{http://www.ccs.neu.edu/home/amislove/twittermood/}   
study \cite{BieverC2010Twitter}. 
Each tweet was rated ten times to get a more reliable estimate of
the human-perceived mood, and each rating was a sentiment strength with an
integer between 1 (negative) and 9 (positive).
The average over the ten values represented the canonical ``ground
truth'' for this study.
The tweets were not used during the construction of the word list. 

To compute a sentiment score of a tweet I identified words and
found the valence for each word by lookup in the sentiment lexicons.
The sum of the valences of the words divided by the number of words
represented the combined sentiment strength for a tweet. 
I also tried a few other weighting schemes: The sum of valence
without normalization of words, normalizing the sum with the number of
words with non-zero valence, choosing the most extreme valence
among the words and quantisizing the tweet valences to +1, 0 and
--1. For ANEW I also applied a version with match using the NLTK
WordNet lemmatizer. 

\section{Results}

\begin{wrapfigure}[15]{r}{0.5\textwidth}
  \vspace{-6em}
  \centering
  \includegraphics[width=.5\textwidth]{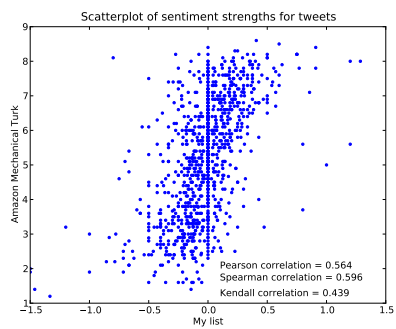}
  \caption{Scatter plot of sentiment strengths for 1,000 tweets with
    AMT sentiment plotted against sentiment found by application or my
    word list.} 
  \label{fig:tweetscatter}
\end{wrapfigure}

My word tokenization identified 15,768 words in total among the 1,000
tweets with 4,095 unique words. 
422 of these 4,095 words hit my 2,477 word sized list, while the
corresponding number for ANEW was 398 of its 1034 words.
Of the 3392 words in General Inquirer I labeled with non-zero
sentiment 358 were found in our Twitter corpus and for OpinionFinder
this number was 562 from a total of 6442.

\begin{wraptable}{r}{.5\textwidth}
  \vspace{.5em}
  \begin{center}
    \begin{tabular}{c|ccccc}
           & \makebox[9mm]{My} &  \makebox[9mm]{ANEW} 
           & \makebox[9mm]{GI}  & \makebox[9mm]{OF} & \makebox[9mm]{SS} \\
      \hline
      AMT  & .564 & .525 & .374 & .458 & .610  \\
      My  &      & .696 & .525 & .675 & .604 \\
      ANEW &      &      & .592 & .624 & .546 \\
      GI   &      &      &      & .705 & .474 \\
      OF   &      &      &      &      & .512
    \end{tabular}
    \caption{Pearson correlations between sentiment strength
      detections methods on 1,000 tweets. AMT: Amazon Mechanical Turk,
      GI: General Inquirer, OF:
      OpinionFinder, SS: SentiStrength.} 
    \label{tab:corrmatrix}
  \end{center}
\end{wraptable}

I found my list to have a higher correlation (Pearson correlation:
0.564, Spearman's rank correlation: 0.596, see the scatter plot in
Figure~\ref{fig:tweetscatter}) with the labeling
from the AMT than ANEW had (Pearson: 0.525,
Spearman: 0.544).
In my application of the General Inquirer word list it did not
perform well having a considerable lower AMT correlation than my list
and ANEW (Pearson: 0.374, Spearman: 0.422).
OpinionFinder with its 90\% larger lexicon performed better than
General Inquirer but not as good as my list and ANEW (Pearson: 0.458, Spearman: 0.491). 
The SentiStrength analyzer showed superior performance with a Pearson
correlation on 0.610 and Spearman on 0.616, see 
Table~\ref{tab:corrmatrix}.

I saw little effect of the different tweet sentiment scoring
approaches: For ANEW 4 different Pearson correlations were in the range
0.522--0.526.
For my list I observed correlations in the range 0.543--0.581 with
the extreme scoring as the lowest and sum scoring without
normalization the highest. 
With quantization of the tweet scores to +1, 0 and --1 the
correlation only dropped to 0.548. 
For the Spearman correlation the sum scoring with normalization for
the number of words appeared as the one with the highest value
(0.596).

\begin{wrapfigure}[20]{r}{.5\textwidth}
  \vspace{-1em}
  \centering
  \includegraphics[width=.5\textwidth]{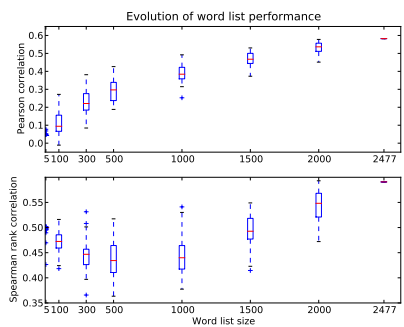}
  \caption{Evolution of performance as the word list is extended with
    from 5 words to the full set of words (2477). The upper panel is
    for the Pearson correlation while the lower for the Spearman rank
    correlation. The boxplots are generated from 50 resamples among the 2477 words.} 
  \label{fig:evolution}
\end{wrapfigure}

Figure~\ref{fig:evolution} plots the evolution of the performance of
the word list on the Twitter as the word list is extended from 5
words to the full set of 2477 words. 

To examine whether the difference in performance between the
application of ANEW and my list is due to a different lexicon or a
different scoring I looked on the intersection between the two word
lists.  
With a direct match this intersection consisted of 299 words. 
Building two new sentiment lexicons with these 299 words, one with the
valences from my list, the other with valences from ANEW, and
applying them on the Twitter data I found that the Pearson
correlations were 0.49 and 0.52 to ANEW's advantage.

\section{Discussion}

On the simple word list approach for sentiment analysis I found my
list performing slightly ahead of ANEW.  
However the more elaborate sentiment analysis in SentiStrength showed
the overall best performance with a correlation to 
AMT labels on 0.610. 
This figure is close to the correlations reported in the evaluation of
the Senti\-Strength algorithm on 1,041 {MySpace} 
comments (0.60 and 0.56)  \cite{ThelwallM2010Sentiment}.

Even though General Inquirer and OpinionFinder have the largest word
lists I found I could not make them perform as good as
SentiStrength, my list and ANEW for sentiment strength detection in
microblog posting.  
The two former lists both score words on polarity rather than
strength and it could explain the difference in performance.

Is the difference between my list and ANEW due to better scoring or
more words? 
The analysis of the intersection between the two word list indicated
that the ANEW scoring is better. 
The slightly better performance of my list with the entire lexicon
may be due to its inclusion of Internet slang and obscene words. 

Newer methods, e.g., as implemented in SentiStrength, use a range of
techniques: detection of negation, handling of emoticons and spelling
variations \cite{ThelwallM2010Sentiment}.
The present application of my list used none of these approaches and
might have benefited. 
However, the SentiStrength evaluation showed that valence switching at
negation and emoticon detection might not necessarily increase the
performance of sentiment analyzers (Tables 4 and 5 in
\cite{ThelwallM2010Sentiment}). 

The evolution of the performance (Figure~\ref{fig:evolution})
suggests that the addition of words to my list might still
improve its performance slightly. 

Although my list comes slightly ahead of ANEW in Twitter sentiment
analysis, ANEW is still preferable for scientific psycholinguistic
studies as the scoring has been validated across several persons.   
Also note that ANEW's standard deviation was not used in the
scoring. It might have improved its performance.

\section*{Acknowledgment}

I am grateful to Alan Mislove and Sune Lehmann for providing the 1,000
tweets with the Amazon Mechanical Turk labels and to Steven J. DeRose
and Greg Siegle for providing their word lists.
Mislove, Lehmann and Daniela Balslev also provided input to the article.
I thank the Danish Strategic Research Councils
for generous support to the `Responsible Business in the
Blogosphere' project.

\bibliographystyle{splncs}

\end{document}